\documentclass[]{aa}  
\pdfoutput=1
\usepackage{graphicx}
\usepackage{epsf}
\usepackage{longtable,lscape}
\usepackage{dcolumn}
\usepackage{booktabs} 
\usepackage{siunitx}
\usepackage{url}
\usepackage[varg]{txfonts}
\usepackage{natbib,twoopt} 

\bibpunct{(}{)}{;}{a}{}{,} 
\newcommand{\bphi}{\mbox{\boldmath $\phi$}}
\newcommand{\bomega}{\mbox{\boldmath $\omega$}}

\let\oldhat\hat
\renewcommand{\vec}[1]{\mathbf{#1}}
\renewcommand{\hat}[1]{\oldhat{\mathbf{#1}}}

\begin{document}

   \title{The impact of rotation on the line profiles of Wolf-Rayet stars}

   \subtitle{}

   \author{T. Shenar\inst{1}
          \and W.-R. Hamann\inst{1}
          \and H. Todt\inst{1}
          }

   \institute{\inst{1} Institut f\"ur Physik und Astronomie, Universit\"at Potsdam,
              Karl-Liebknecht-Str. 24/25, D-14476 Potsdam, Germany\\
              \email{shtomer@astro.physik.uni-potsdam.de}
              }
   \date{Received August 16, 2013 / Accepted January 8, 2014}


\abstract
{Massive Wolf-Rayet stars are recognized today to be in a very common, but short, evolutionary phase of massive stars.
While our understanding of Wolf-Rayet stars has increased
dramatically over the past decades, it remains unclear whether rapid rotators are among them. 
There are various indications that  
rapidly rotating Wolf-Rayet stars should exist. Unfortunately, due to their expanding atmospheres,
rotational velocities of Wolf-Rayet stars are very difficult to measure.
However, recently observed spectra of several Wolf-Rayet stars reveal peculiarly broad
and round emission lines. 
Could these spectra imply rapid rotation?}
{In this work, we model the effects of rotation on the atmospheres of
    Wolf-Rayet stars. We further investigate whether the peculiar spectra of five Wolf-Rayet stars may be
explained with the help of stellar rotation, infer appropriate rotation parameters, 
and discuss the implications of our results.}
{We make use of the Potsdam Wolf-Rayet (PoWR) non-LTE model atmosphere code. 
Since the observed spectra of Wolf-Rayet stars are mainly formed in 
their expanding atmospheres, 
rotation must be accounted for 
with a 3D integration scheme of the formal integral. For this purpose, 
we assume a rotational velocity field consisting of an inner co-rotating domain and an outer domain,
where the angular momentum is conserved.}
{We find that rotation can reproduce the unique spectra analyzed here. However,
       the inferred rotational velocities at the stellar surface are large ($\sim200$\,km/s), and the 
       inferred co-rotation radii ($\sim10 R_*$) suggest the existence of very strong photospheric magnetic fields ($\sim20\,$kG).}
{}

\maketitle

\section{Introduction}
\label{sec:introduction}
All massive stars at solar metallicity with initial masses larger than~$\sim20 M_\odot$ eventually become 
Wolf-Rayet (WR) stars \citep{Sander2012}. WR stars have a tremendous influence on their host galaxies, owing to 
their ionizing radiation and the immense transfer
of kinetic energy and momentum to the surrounding interstellar matter, carried with their powerful stellar winds.
Their study is thus fundamental  
to our understanding of the evolution of massive stars, and as a consequence, of
galaxies 
\cite[cf.][for a thorough review on the properties of WR stars]{Crowther2007}. 

Knowing whether WR stars might exhibit rapid rotation is 
crucial in this context. Rotation dramatically affects the 
chemical stratification of the 
star due to
rotationally induced mixing \cite[e.g.][]{Heger2000}, may cause surface 
deformations such as gravity 
darkening \citep{VonZeipel1924} or disk formation \citep{Bjorkman1993}, and may also contribute 
to the driving of the wind \citep{Friend1986}.   

Various theoretical considerations and models suggest that rapidly rotating WR stars may exist.    
It is observed that many O stars, which are the direct progenitors of WR stars, exhibit
rotational velocities of several hundred km/s \cite[cf.][]{Penny1996, Vanbeveren1998}.
It has been recently suggested \cite[e.g.][]{DeMink2013} that the highest rotational velocities 
observed in O stars might be the result 
of binary interaction or even merging processes.
Indeed, 
the stellar wind severely dampens the rapid rotation of the star shortly before the WR phase initiates.
On the other hand, the continuous contraction of the star leads to 
faster rotation, so that the stellar rotation of a WR star is determined
by these two counteracting mechanisms.
Evolutionary models calculated by 
\cite{Meynet2005}  
with initial rotational velocities 
of $300$\,km/s 
predict, especially at low metalliticies, a considerable \emph{increase} of the rotational velocity as the 
WR star evolves from a late subtype (WNL) to an early subtype (WNE),
reaching values of up to $200$\,km/s. 
WR stars with rapidly rotating cores have been proposed 
as progenitors of long-duration gamma-ray bursts (LGRBs) 
by various authors \cite[e.g.][]{Vink2005, Hirschi2005, Woosley2006, Graefener2012}, relying on the collapsar model of \cite{Woosley1993_2}.

Usually, rotational velocities of stars are determined from the rotational broadening of their absorption lines. However,
this method can hardly be applied to WR stars.
The spectra of WR stars, especially of early subtypes, are predominantly formed in their dense, 
extended winds, and thus rarely exhibit such absorption lines in their observed spectra.

Various methods to probe the rotation of WR stars have been
proposed throughout the years. A few WR stars have been observed to show
periodic photometric variations \cite[e.g.][]{Marchenko1998} which may be
attributated to so called 
co-rotating interaction regions (CIRs) in the wind \cite[][]{Owocki1996, Chene2005}.
Rapid rotation may also lead to an axisymmetric
density structure \cite[e.g.][]{Coyne1982, Ignace1996}.
If layers in which electron scattering occurs become axisymmetric, 
the departure from spherical symmetry may be detected with linear polarimetry
via the so-called line effect \cite[e.g.][]{Harries1998}, i.e.\ an enhanced polarization of the 
continuum radiation 
relative to that of emission lines. 
While a positive detection of the line effect may be associated with rapid rotation, 
other scenarios such as close binaries \cite[][]{Brown1977}
or strong magnetic fields \citep[][]{Poe1989} were shown to also cause it. Moreover, 
rotational velocities are very hard to quantify on the basis
of polarimetric measurements alone.

Recent spectroscopic studies
performed with the non-LTE Potsdam Wolf-Rayet (PoWR) model atmosphere code
potentially imply the existence of rapidly rotating WR stars.
\cite{Hamann2006} performed a flux-convolution with a rotation profile \cite[e.g.][]{Unsold}
in order to reproduce the broad and round emission lines of the Galactic 
WN2 star WR 2. Hainich et al.\ (submitted), who extensively analyzed 102 WN stars residing in the Large Magellanic Cloud
(LMC),
also made use of flux-convolution with rotation profiles of very large rotational velocities for a few of them.
\cite{Sander2012} used the same technique
for two WO stars. So far, at least
ten WR stars have been identified to exhibit round and broad line profiles, and as many as half of them show 
no signs for binarity.

Alas, the method of flux-convolution with rotation profiles is not valid in the case of expanding atmospheres. 
The emergent radiation is not simply emitted from one single rotating layer, but rather formed in the
expanding stellar wind, a region that extends over tens of stellar radii. 
To account for rotation in the case of expanding atmospheres,  
a model of the rotational velocity field is prerequisite. 
The evaluation of the formal integral must 
include the effects of the modified velocity field,  
which is comprised of 
radial and rotational components. 
Nevertheless, the fact that rotation profiles help to reproduce
the observed spectra of several WR stars is potentially a beacon for rapid rotation in these stars. 

\begin{figure}[htb]
\centering
  \includegraphics[width = \hsize]{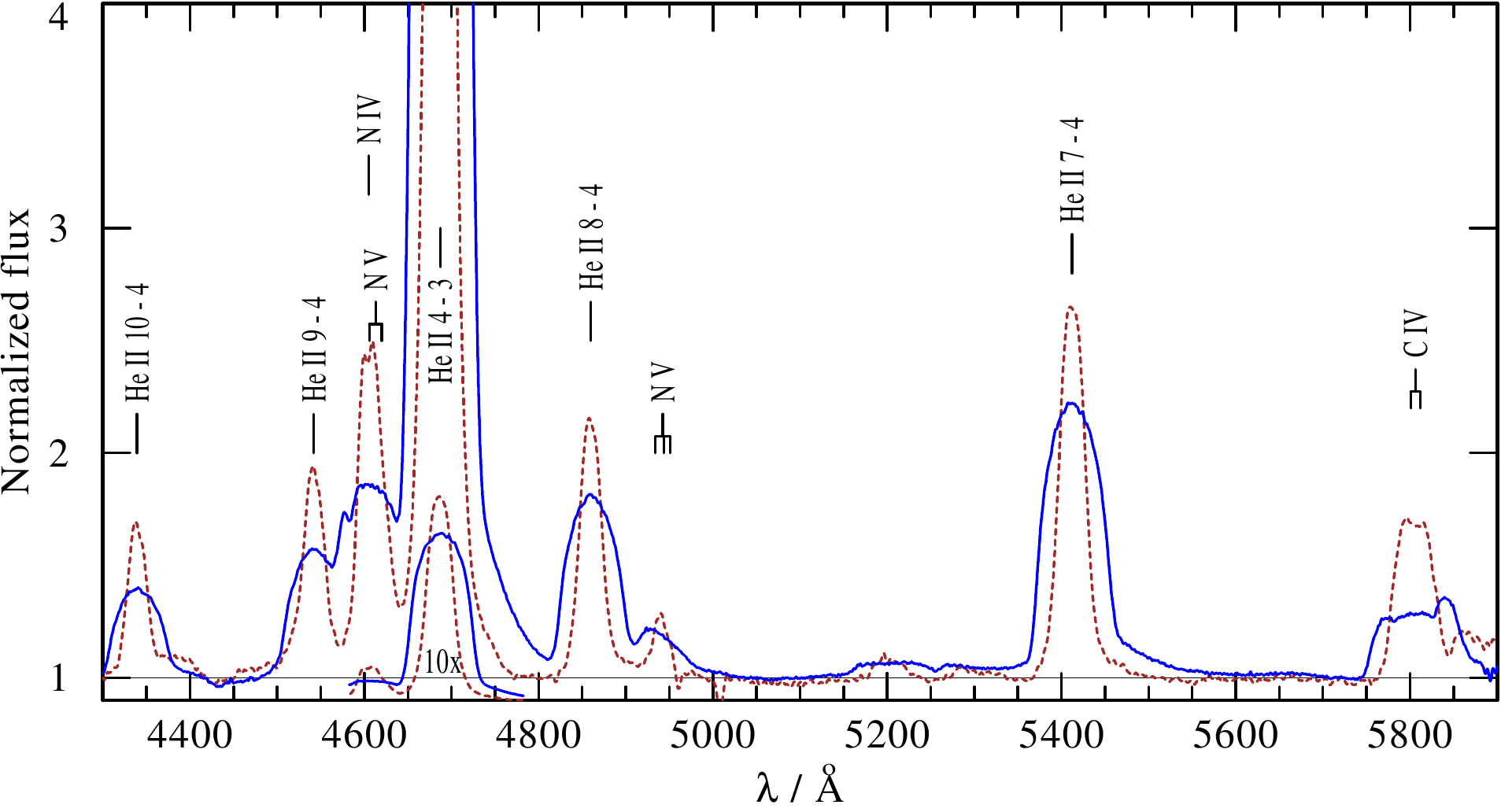}
  \caption{Comparison between the observed optical spectra of two WN4b stars in the LMC: BAT99 134 (dashed brown line),
and BAT99 7 (solid blue line). While the spectrum of BAT99 134 is typical for a WNE subtype, 
the spectrum of BAT99 7 distinctly shows qualitative differences.}
\label{fig:spitzcomp}
\end{figure}

To emphasize the striking uniqueness of the spectra analyzed here, we show a comparison (Fig.\,\ref{fig:spitzcomp})
between
the observed spectra of two WN4b stars residing in the LMC:
BAT99 7 (solid blue line), analyzed in this work for rapid rotation, and 
\mbox{BAT99 134} (dashed brown line). 
Note the distinct qualitative difference between the two spectra --
and this despite their identical spectral class! As opposed to the sharp, triangular-like emission lines of \mbox{BAT99 134}, which are very typical for 
WNE stars,
the spectrum of \mbox{BAT99 7} exhibits extremely round and broad spectral lines. Note also how the C\,{\sc iv}\,$\lambda5800$ multiplet appears 
stretched and flat-topped in the spectrum of BAT99 7.
Out of hundreds of WR stars previously analyzed, only a handful exhibit such exceptional features. 
  
\cite{Ignace2009} managed to obtain round and broad line profiles as an analytical solution 
of the transfer equation in the case of optically
thick lines and for a specific parameter selection. However, his assumptions (e.g.\ 
constant wind velocity, LTE population numbers) do not  
hold in the case of WR stars. Most importantly, such line profiles are not obtained
using non-LTE spectral modeling, independently of the parameter choice. 
One might further argue that round emission lines could be reproduced by assuming much larger terminal velocities. While
a larger terminal velocity indeed yields broader lines, it does not help reproduce their round profiles, 
as we later show. 
We therefore currently see no 
alternatives which can explain the unique spectra analyzed here other than rotation.
The prediction that rapidly rotating WR~stars may exist, along with the existence of such peculiar spectra, provide 
the motivation for a proper treatment of rotation in expanding atmospheres of WR~stars.
 
Our paper is structured as follows: Section\,\ref{sec:obsdat} consists of a short description of our
sample and of the corresponding observational data. In Sect. \ref{sec:model}, we briefly describe 
the model atmosphere code and the rotational velocity field. In Sect.\,\ref{sec:analysis}, we explain our
methods in detail and present our results.
Finally, in Sect.\,\ref{sec:discussion}, we thoroughly discuss the plausibility and implications of our results.

\section{The sample and the observational data}
\label{sec:obsdat}

We analyze five stars in this work, comprising four LMC stars (BAT99 7, 51, 88, and 94), 
and one Galactic star (WR 2). These are the only 
stars, out of the $\sim 180$ WN stars analyzed by \cite{Hamann2006} and Hainich et al.\,(submitted), 
which exhibit peculiarly broad and round spectral emission lines and are not classified
as binaries. Stars which were classified as binaries in previous studies \cite[e.g.][]{Schnurr2008, Foellmi2003}, or 
whose spectrum appears to be a composite of two spectra, are omitted from this analysis; 
their study is postponed to future work. 
No WC stars out of those analyzed by \cite{Sander2012} show round lines in their observed spectra.
To restrict the discussion to WN stars, the analysis of WO stars is not included in this work.  

The observation used for WR 2 were taken 
at the ``Deutsch-Spanisches Astronomisches Zentrum (DSAZ)'', Calar Alto, Spain with the Boller \& Chivens Cassegrain
spectograph at the 2.2m-telescope. The spectrum covers a spectral range of 3320 to $7400\,\AA$ 
with a spectral resolution of $\approx3600$ \cite[cf.][for more details]{Hamann1995}.
The optical spectrum of the LMC star BAT99 7 in the domain $4000-6000\,\AA$  was taken 
on 2010 December 23 with FORS2 at the VLT (PI: W-R.\ Hamann) with a spectral resolution of $\approx1600$. For the spectral domain $6000-6800\,\AA$
of the star BAT99 7 and for the remaining LMC stars analyzed here, we
use co-added, rectified spectra, kindly provided   
by C. Foellmi.  The individual spectra were 
taken with various instruments, as described by \cite{Foellmi2003}\footnote{The single spectra are available at
\url{http://wikimbad.obs.ujf-grenoble.fr/Category_Wolf-Rayet_Star.html}}, with a spectral resolution of about 1000.

\section{The model}
\label{sec:model}

\subsection{PoWR Model atmospheres}
\label{subsec:PoWR}
The PoWR model atmosphere code solves the radiative transfer and rate 
equations in the co-moving frame, calculates the non-LTE population numbers,
and delivers a synthetic spectrum by evaluating the formal integral 
in the observer's frame.
A closer description 
of the assumptions (spherical symmetry, stationary mass-loss) 
and methods used in the code is given by \cite{Graefener2002} and
\cite{Hamann2004}. 
Clumping in the wind is accounted for by the use of a constant clumping
factor $D$ which describes the density ratio of a clumped wind to an identical homogenous one \cite[cf.][]{Hamann1998}. 
Line blanketing is treated using the superlevel approach \cite[][]{Graefener2002}, as 
originally  
implemented by \cite{Hillier1998}.

The inner boundary of a model is denoted with $R_*$. It is defined at the
Rosseland optical depth \mbox{$\tau_\text{Ross} =  20$} \cite[cf.][]{Hamann2006}.
For a given luminosity $L_*$, the stellar temperature
$T_*$ is defined via the Stefan-Boltzmann~relation. 

We do not calculate hydrodynamically consistent model atmospheres in this work, but rather use a pre-specified 
wind velocity law. In the subsonic region,
the velocity field is defined such that a hydrostatic density stratification is approached.  
In the supersonic region, the radial wind velocity at a distance $r$ from the center 
takes the form of the so-called $\beta$-law:
\begin{equation}
 v_r (r) = v_\infty \left( 1 - \frac{R_*}{r} \right)^\beta
\label{eq:beta}
\end{equation}

The exponent $\beta$ is fixed to the standard value $\beta =1$ for all models calculated here. The
influence of different values of $\beta$ is discussed in Sect.\,\ref{subsec:Results}.
The terminal velocity $v_\infty$ is determined individually
for each star analyzed here. For the Doppler velocity $v_\text{D}$ corresponding to 
microturbulent and thermal motion, we adopt the constant value $v_\text{D} = 100$\,km/s
\citep[cf.][]{Hamann2000}.

Once the population numbers in the wind have been established,
the emergent intensities are calculated by solving the transfer equation along rays
parallel to the line-of-sight (the formal integral).
Without rotation, it is sufficient to evaluate the 
emergent 
intensities at distinct impact 
parameter values only, and finally to integrate over all 
impact parameters. The azimuthal integration is 
in this case redundant. 

\begin{figure}[htb]
  \centering
  \includegraphics[width = \hsize]{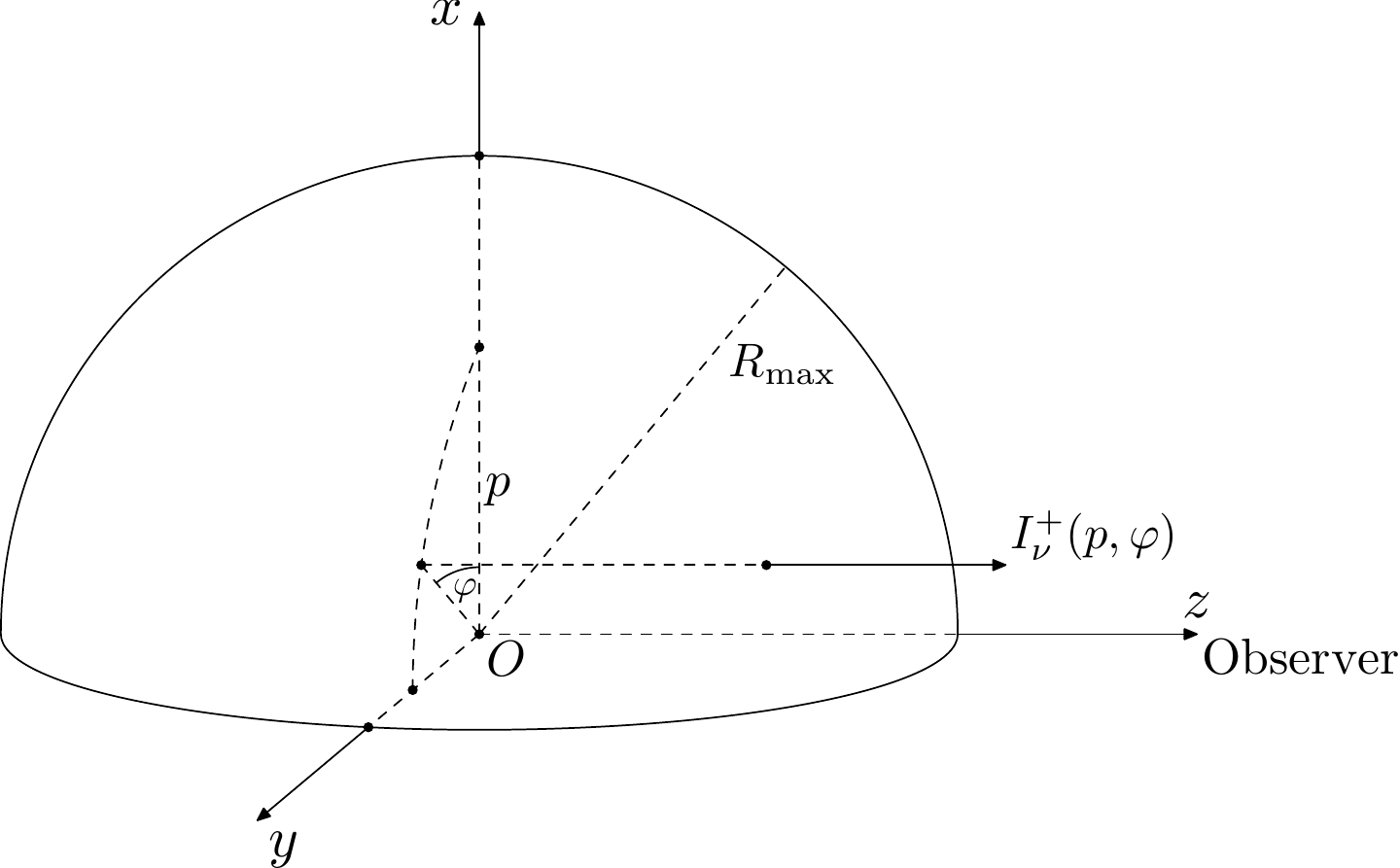}
  \caption{The integration of the emergent intensities $I^+_\nu(p,\varphi)$ has to be performed over 
           the impact parameter $p$ and the azimuthal angle $\varphi$. The $z$-axis in the 
           figure points towards the observer. The $x$- and $y$- axes form together with the $z$-axis a 
           Cartesian coordinate system.}
\label{fig:emint}
\end{figure}

If spectral lines are formed in a non-extended stellar photosphere, 
the azimuthal integration can be avoided even in the 
presence of rotation. 
In this case, the effect of rotation may be accounted for very accurately by performing a flux-convolution 
with 
rotation profiles \cite[e.g.][]{Unsold}. However, since the spectra of WR stars are formed in their 
dense and extended winds, 
the use of rotation profiles is not justified. Rather, for the calculation of the flux $f_\nu$ 
at
a distance $d$ from the star, 
the emergent intensities $I^+_\nu(p,\varphi)$ are to be 
integrated over all impact parameters $p$ and all azimuthal 
angles~$\varphi$:

\begin{equation}
 f_\nu = \frac{1}{d^2}\int\limits_{\varphi=0}^{2\pi}\int\limits_{p=0}^{R_\text{max}}\, I_\nu^+(p,\varphi)\, p \, 
 \mathrm{d} p \, \mathrm{d} \varphi 
\label{eq:specstar}
\end{equation}
where $R_\text{max}$ is the upper integration boundary which includes the entire wind (see Fig.\,\ref{fig:emint}). 

The emergent intensity $I^+_\nu(p,\varphi)$ is obtained by integrating the transfer equation 
over the optical depth $\tau$ along the ray specified by $p$ and $\varphi$.
This evaluation of the formal integral over $\tau$, together with the integration of $I^+_\nu(p,\varphi)$ over $p$ and $\varphi$, covers 
the wind in all three dimensions, and hence we name this calculation the ``3D integration scheme''.

\subsection{The rotational velocity field}
\label{subsec:rotmodel}
The rotational velocity field adopted here relies on a few assumptions. 
The stellar atmosphere (including the wind) is divided into two radial domains: 

\begin{enumerate}
\item The co-rotating domain $r \le R_\text{cor}$, where we postulate the existence of a radius $R_\text{cor} \geq R_*$ up to which the 
      wind co-rotates with the star
      with the \mbox{angular velocity $\vec{\bomega}$}.
\item The outer domain $r > R_\text{cor}$, where the only forces present are all radial, and where  
      the angular momentum is therefore conserved.
\end{enumerate}
 
We thoroughly discuss the motivation behind these assumptions in Sect.\,\ref{sec:discussion}.
As a first order approximation, we account for rotation only in the formal integral. 
\cite{Petrenz2000} discuss possible effects of rotation on the population numbers in early-type stars.
For simplicity, we ignore effects caused by deformations of the stellar surface such as gravity darkening
and disk formation. 
The angular velocity $\vec{\bomega}$ of the co-rotating domain ($r \le R_\text{cor}$)
is assumed to be independent of the latitude~$\theta$. 
Based on hydrodynamical studies by \cite{Owocki1994}, we neglect  
the polar components of the velocity field, which are much smaller than the
rotational and radial components.

These assumptions allow one to deduce the form of 
the rotational velocity field on the basis of simple 
physical arguments. Denoting with $r$ the distance between a wind element and the center of the star, 
with $\hat{\bphi}$ the azimuthal unit vector relative to the rotation axis $\vec{\bomega}$,
and with $\theta$ the latitude (measured from the positive direction of the rotation axis), one finds the rotational velocity field
$v_\phi$ to be:
\begin{equation}
 v_\phi(r,\theta) = v_\text{eq} \left(R_\text{cor}\right)  \sin \theta
\left\{
    \begin{array}{cl}\displaystyle
                \frac{r}{R_\text{cor}}                         &  r \leq  R_\text{cor} \\[1em] 
                \displaystyle
                \frac{R_\text{cor}}{r}  & r > R_\text{cor}
    \end{array}
\right.
\label{eq:phiform2}
\end{equation}
where $v_\text{eq}\left(R_\text{cor}\right)$ is the equatorial rotational velocity \emph{at the co-rotation radius}.

Since we observe the star at an unknown inclination \mbox{angle $i$} (the angle between the line-of-sight
and the rotation axis $\vec{\bomega}$), 
the only parameter having a direct observable influence on the spectrum is $v_\text{eq}\left(R_\text{cor}\right) \sin i$, denoted as of now simply 
with $v_\text{cor} \sin i$. 
$R_\text{cor}$ and $v_\text{cor} \sin i$ fully define the rotational velocity field, and are 
regarded as free parameters in this analysis.

\begin{table*}[htb]
\caption{Parameters of the suspected rapid rotators} 
\begin{center}
\small
\begin{tabular}{llllll}
 \toprule
     Object  & \object{BAT99 7}&\object{BAT99 51}&\object{BAT99 88}&\object{BAT99 94}& \object{WR 2} \\
 \midrule
  Spectral type &  WN4b & WN3b & WN4b/WCE & WN4b & WN2-w\\[0.3mm]
  $T_*$\,[kK]& 141 & 89 & 112 & 141 & 141\\[0.3mm]
  $\log R_\text{t}$\,[$R_\odot$]& 0.1 & 0.6 & 0.4 & 0.0 & 0.5 \\[0.3mm]
  $R_*$\,[$R_\odot$]& 1.3 & 1.9 & 2.1 & 1.3 & 0.9\\[0.3mm]
  $\log \dot{M}$\,[$M_\odot\,\text{yr}^{-1}$]& -4.5 & -5.2 & -4.8 & -4.4& -5.3\\[0.3mm]
  $\log L$\,[$L_\odot$]& 5.84 & 5.30 & 5.80 & 5.80 & 5.45 \\[0.3mm]
  $M$\,[$M_\odot$]& 26 & 13 & 25 & 25 & 16\\ [0.3mm]
  $\eta_\text{mom}$ &  5.5 & 2.5 & 2.0 & 6.2 & 1.6\\[0.3mm]
  $X_\text{N}$ & $0.015^\text{(a)}$ & 0.01 & 0.008 & 0.01 & 0.001 \\[0.3mm]
  $X_\text{C}$ & $7\cdot10^{-5}$ & $7\cdot10^{-5}$ & $0.005^\text{(a)}$ & $7\cdot10^{-5}$ & $1\cdot10^{-4}$  \\
  \midrule
  $v_\infty$\,[km/s]& 2400 & 1600 & 1600 & 2000 & 1800\\   [0.3mm]
  $v_\text{cor} \sin i$\,[km/s]& $2500\pm500$ & $2000\pm400$ & $1700\pm340$ & $2500\pm500$ & $3000\pm600$\\ [0.3mm] 
  $R_\text{cor}$\,[$R_*$]& $16\pm5$ & $10\pm3$ & $12\pm4$ & $12\pm4$ & $6\pm2$\\ [0.3mm]  
  $v_* \sin i$\,[km/s]& $160\pm80$ & $200\pm100$ & $140\pm70$ & $210\pm100$ & $500\pm250$ \\
\bottomrule
 \end{tabular}
\label{tab:rotstars}
\end{center} {\footnotesize
The stellar parameters not involving the velocity field (upper part) were primarily adopted from 
Hainich et al.\ (submitted) for the LMC stars, and from \cite{Hamann2006} for WR 2. All sampled stars 
are calculated with Fe mass fractions of 
$X_\text{Fe} = 7\cdot10^{-4}$. The LMC stars are calculated with a clumping factor of $D = 10$, while $D=4$ was
adopted for WR~2.
The uncertainties are discussed in Sect.\,\ref{subsec:Errors}.\\
(a) The high abundances were needed to obtain the observed strength of the exceptionally 
strong nitrogen/carbon spectral lines.}
\end{table*}

$v_\text{cor} \sin i$ 
is not to be confused with the stellar rotational velocity $v_* \sin i$, defined as the rotational velocity
at $r = R_*$. 
Since the velocity increases linearly in the rigid-body
regime, the rotational velocity at $r = R_*$ is given by
\begin{equation}
 v_*\sin i = \frac{v_\text{cor} \sin i}{R_\text{cor} / R_*} 
\end{equation}

It is thus important to keep in mind that $v_* \sin i$ values could easily be an order of magnitude smaller than 
corresponding $v_\text{cor} \sin i$ values. 
$v_* \sin i$ is the velocity one usually refers to when discussing the stellar rotational velocity.

The extended version of PoWR was tested on several stars which are known to rotate. 
For stars with photospheric lines,
the new version was shown to yield almost identical synthetic spectra to those obtained by means of flux-convolution with a 
suitable rotation profile. Furthermore, the code was tested on the O 
supergiant $\zeta$ Pup. Similarly to \cite{Hillier2012}, 
we find that rotation, accounted for with the 3D integration scheme,
reproduces the observed emission lines much better than a simple flux-convolution.

\section{Analysis}
\label{sec:analysis}
\subsection{Stellar parameters}
\label{subsec:stelpam}

The stellar parameters used for the calculation of the synthetic spectra are summarized
in Table\,\ref{tab:rotstars}.
The basic stellar parameters are listed at the upper part of the table.
The values are  
based on previous studies performed by Hainich et al.\ (submitted) and \cite{Hamann2006}, though
slight modifications were performed due to 
the inclusion of rotation in the present studies. The velocity 
parameters inferred in this work
are listed at the bottom part
of Table\,\ref{tab:rotstars}. The method of their determination is discussed 
in Sect.\,\ref{subsec:rotpardet}.

$R_\text{t}$, given in Table\,\ref{tab:rotstars}, is the so called ``transformed radius'', originally 
introduced by \cite{Schmutz1989}, and later generalized to include the effect of 
clumping by \cite{Hamann1998}.
\cite{Schmutz1989} noticed that
emission lines belonging to models calculated with identical temperatures $T_*$, chemical abundances,  
and $R_\text{t}$ values, have similar equivalent widths, independent of the
mass-loss rate $\dot{M}$, radius $R_*$, and \mbox{terminal velocity $v_\infty$}.   

The wind efficiency $\eta_\text{mom}$, which is given in Table\,\ref{tab:rotstars},
is defined as the ratio
of the wind momentum to the radiation momentum
\begin{equation}
 \eta_\text{mom} := \frac{\dot{M} v_\infty}{L / c}
\end{equation}

If the wind is driven only by radiation, $\eta_\text{mom}$
corresponds to the average number of scatterings in the wind per photon.

\subsection{Parameter study}
\label{subsec:parstudy}

\begin{figure*}[htb]
 \centering
 \includegraphics[width = \hsize]{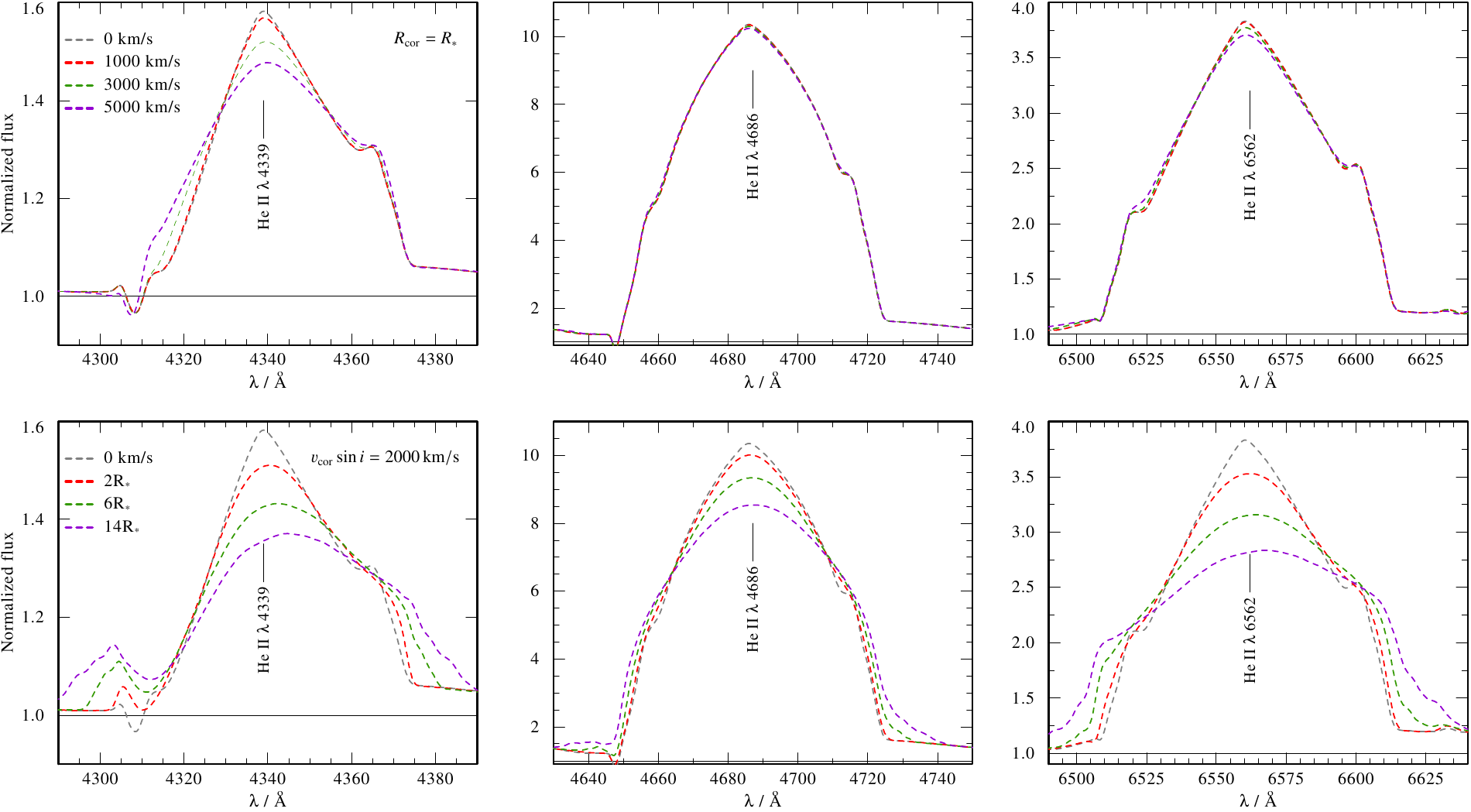}
 \caption{The effect of the parameters $v_\text{cor} \sin i$ and $R_\text{cor}$ on prominent synthetic emission lines of BAT99 7. 
          Each of the upper plots depicts, for a constant value $R_\text{cor} = R_*$, the synthetic profile
          of an emission line calculated with different $v_\text{cor} \sin i$ values. 
          The bottom plots show the synthetic profile of the same spectral lines 
          calculated with different $R_\text{cor}$ values and with a constant rotational velocity of $v_\text{cor} \sin i = 2000$\,km/s.}
\label{fig:partest}
\end{figure*}

To demonstrate how the synthetic spectra of WR stars respond to the rotation parameters $v_\text{cor} \sin i$ and $R_\text{cor}$,
we focus on the WN4 star BAT99 7 residing in the LMC. The model parameters are
summarized in Table\,\ref{tab:rotstars}. 
Figure \ref{fig:partest} depicts several synthetic He\,{\sc ii} lines calculated with a constant $R_\text{cor}$ value
and different $v_\text{cor} \sin i$ values (upper panels), 
and the same lines calculated with a constant $v_\text{cor} \sin i$ value and different $R_\text{cor}$ values (bottom panels). 
For the exact values, 
see caption and legend of Fig.\,\ref{fig:partest}. 

As is evident from Fig.\,\ref{fig:partest}, it is only with large co-rotation radii that rotation 
significantly affect the spectrum. It is especially striking how the He\,{\sc ii} $\lambda4686$ line 
hardly responds to the huge rotational velocities applied when $R_\text{cor} = R_*$.
The reason that large $R_\text{cor}$ values are needed to significantly affect the spectra is that 
the line forming regions (LFRs) are situated very far above the stellar surface.

To discuss the radial stratification of the emission strength for the individual emission lines, we make use of the quantity
$\xi(r)$, originally introduced by \cite{Hillier1987}.  
$\xi(r)$ serves as a measure for the energy emitted 
due to a line transition as a function of the distance $r$ from the stellar center.
An expression for $\xi$ is obtained by means of the Sobolev approximation \citep{Castor1970}.
The energy $E$ emitted in the considered line satisfies: 
\begin{equation}
 E \propto \int_{R_*}^{\infty} \xi\,d(\log r)
\label{eq:hilxi}
\end{equation}

\begin{figure}[htb]
 \centering
 \includegraphics[width = 0.35\textheight]{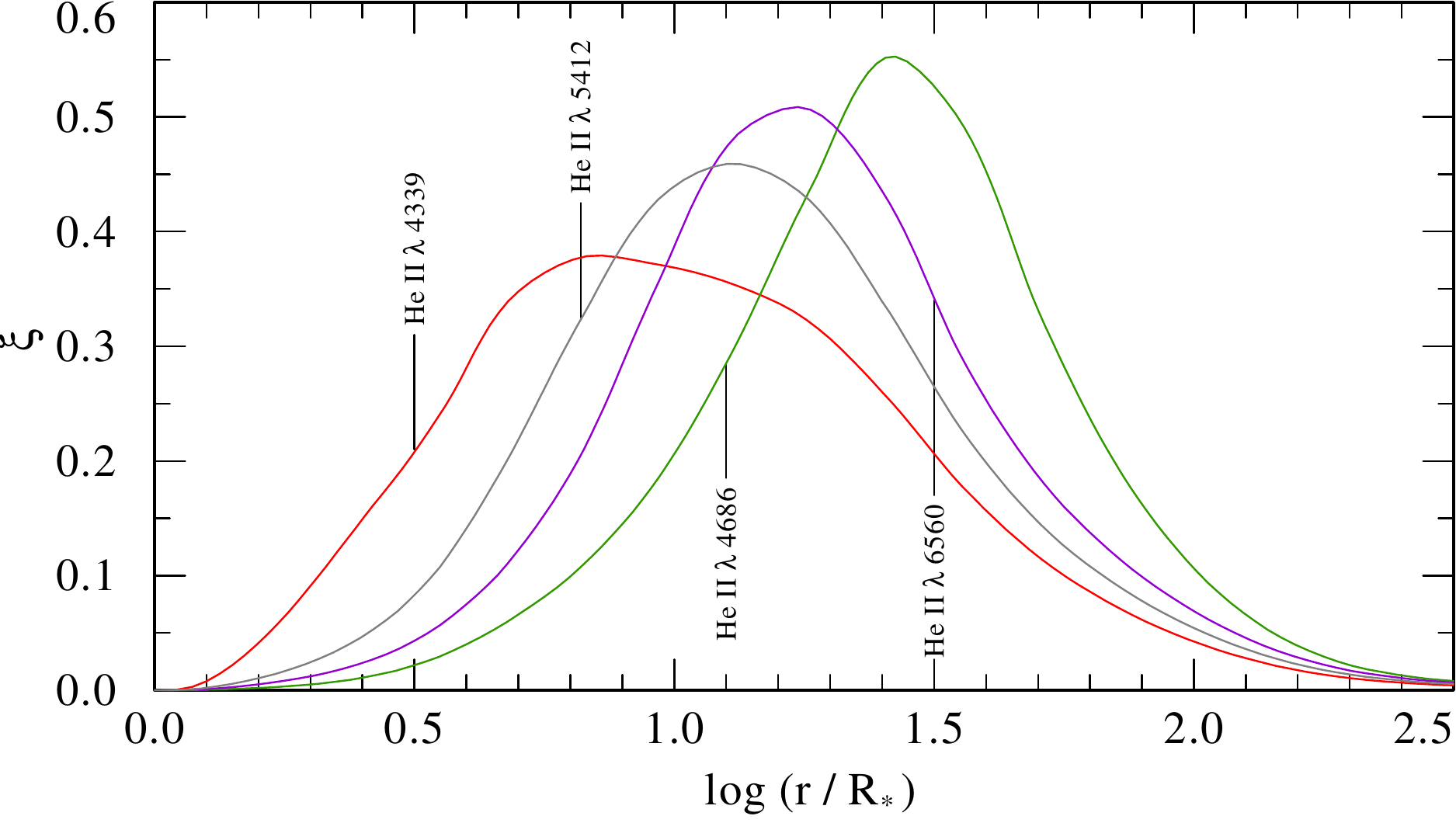}
 \caption{Normalized $\xi(\log r)$-plots for several He\,{\sc ii} emission lines belonging to the model of 
          the star BAT99 7. The model parameters are listed in \mbox{Table\,\ref{tab:rotstars}}. The individual lines are formed 
          many stellar radii above the surface.}
\label{fig:tradwl}
\end{figure}

\begin{figure}[htb]
 \centering
 \includegraphics[width = 0.35\textheight]{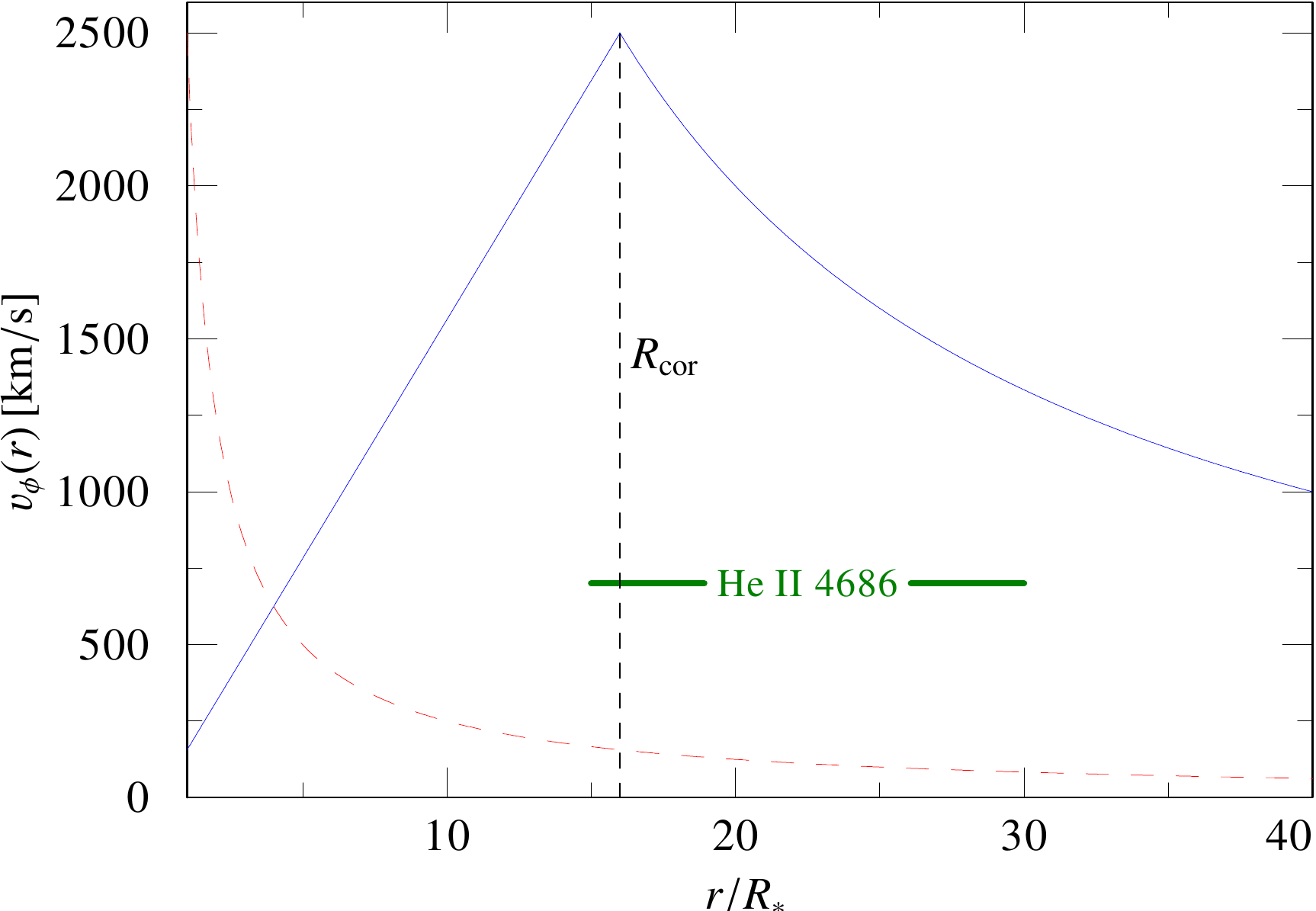}
 \caption{The equatorial rotational velocity as a function of $r$. Both curves correspond to 
          $v_\text{cor} = 2500\,$km/s. However, the dashed red curve is plotted with $R_\text{cor} = R_*$, while the solid blue 
          curve is plotted with $R_\text{cor} = 16 R_*$. We also denote an estimate of the LFR of the line
          He\,{\sc ii}\,$\lambda4686$, determined by its corresponding $\xi$ plot. Note that the LFR remains unaffected 
          by rotation unless large co-rotation radii are assumed.}
\label{fig:rotmodel}
\end{figure}

Fig.\,\ref{fig:tradwl} depicts normalized $\xi(\log r)$-plots for several prominent He\,{\sc ii} emission lines belonging
to the model of BAT99 7, while
Fig.\,\ref{fig:rotmodel} schematically illustrates the rotational velocity as a function of $r$ 
for two cases: $R_\text{cor} = R_*$ (dashed red line) and $R_\text{cor} = 16 R_*$ (solid blue line).
We also indicate in Fig.\,\ref{fig:rotmodel} the
LFR of the line He\,{\sc ii}\,$\lambda4686$, determined by its corresponding $\xi$-plot. 

As discussed in Sect.\,\ref{subsec:rotmodel},
the rotational velocity field consists of two domains: the rigid rotation domain \mbox{$r < R_\text{cor}$}, where the $\phi$-component increases
linearly with $r$, and the outer domain $r > R_\text{cor}$, where the $\phi$-component decreases
with $1/r$.
As can be readily seen, without large co-rotation radii, the rotational velocity (dashed red line) 
falls off rapidly and does not reach significant values in the LFR. 
The same reasoning holds for all prominent lines of the WR stars analyzed here. 
One cannot expect a change of the line profile to occur if the rotational velocity field does not reach significant values
in the corresponding LFR. 

The same analysis was performed with all relevant WNE models,
  and the result remained identical: rotation without co-rotation hardly
  affects the synthetic spectra, independently of the WNE model used. It is thus clear that,
if stellar rotation is to explain the observed spectra, 
\emph{large co-rotation radii have to be assumed}.

\subsection{Determination of the rotation parameters}
\label{subsec:rotpardet}

The determination of the most suitable $v_\text{cor} \sin i$ and $R_\text{cor}$ values is a very difficult task. 
Firstly, different 
combinations of $v_\text{cor} \sin i$ and $R_\text{cor}$ may yield similar line profiles
if they imply a similar rotational velocity in the LFR. Secondly, 
since the inclusion of rotation could affect the equivalent width, previously determined 
stellar parameters may require readjustment when
including rotation.
Lastly, the radial wind velocity $v_r(r)$, characterized by the terminal velocity $v_\infty$,
has a large influence on the width and shape of the lines. 
The value of $v_\infty$ thus becomes coupled to the rotation parameters.

Values of $v_\infty$,
derived using different methods, are given by e.g.\ \cite{Niedzielski2004}, \cite{Niedzielski2002},
\cite{Hamann2000}, and \cite{Prinja1990}. Unfortunately, corresponding $v_\infty$ values often 
differ by $\sim1000$\,km/s between these studies, depending on the spectral line considered.
We thus cannot blindly adopt  values from previous studies for our rotation analysis, which is sensitive 
to the value of $v_\infty$. Instead, the terminal wind velocity $v_\infty$ becomes an additional free parameter 
of our analysis.
Determining the three free parameters 
($v_\text{cor} \sin i$, $R_\text{cor}$, $v_\infty$) without
some guiding strategy can be very tricky, especially due to the computational
cost. 

If we assume that the line width is determined only by Doppler shifts, 
an analytical expression for the maximum line width can be derived. The velocities responsible for the Doppler shift 
are 
the expansion velocity given by the $\beta$-law (Eq.\ \ref{eq:beta}), the rotational velocity 
(Eq.\ \ref{eq:phiform2}), and the Doppler velocity $v_\text{D}$.
Let us assume for a moment that a spectral line $l$ with the wavelength $\lambda_l$ is formed in a relatively well-defined region determined by $r \approx 
r_l$. Because the rotational and expansion velocity vectors depend solely
on $r$ and are always perpendicular to each other, the \emph{maximum} line width due to rotation and expansion is determined 
at the latitude $\theta = i$ and is given by 
$\sqrt{v_r^2(r_l) + v_\phi^2(r_l,i)}$. This width is slightly increased 
due to the much smaller intrinsic thermal and microturbulent motion described by $v_\text{D}$, 
so that the maximum line width is approximately:
\begin{equation}
 \frac{c}{\lambda_l} \Delta \lambda^l_\text{max} = v_{\rm tot}(r_l,i) \cong \sqrt{v_r^2(r_l) + v_\phi^2(r_l,i)} + v_\text{D}
\label{eq:vtotLFR}
\end{equation}

$r_l$ may be very roughly defined as the radius where $\xi_l(r)$ reaches its maximum (cf.\ Fig.\,\ref{fig:tradwl}).  
The observed maximum width of the line $\Delta\lambda^l_\text{max}$ may be measured from a careful inspection of the 
observed spectrum. Having obtained 
a set of data points $(r_l, \Delta\lambda^l_\text{max})$ for a specific star, we can 
approximately 
compare a function of the form of Eq.\ ($\ref{eq:vtotLFR}$) to the data points in the $r-v$ space. 
Of course, the assumption that the lines are formed in a specific radius is an artificial one, so that 
exact parameter values cannot be derived with this method. 
However, this method does help us to find a suitable para\-meter domain of $v_\text{cor} \sin i$, $R_\text{cor}$, and $v_\infty$.

Once a suitable parameter domain has been found, we examine the agreement 
between observations and synthetic spectra calculated with different parameter values in 
the domain.
The final values inferred in this work are 
determined  by identifying the synthetic spectra showing the best agreement with the observations. 
Any uncertainties here (i.e.\ 
the choice of $r_l$)
therefore do not bear an effect on the final values presented in Table\,~\ref{tab:rotstars}.

\subsection{Results}
\label{subsec:Results}

\begin{figure*}[p]
 \centering
  \includegraphics[width = \textwidth]{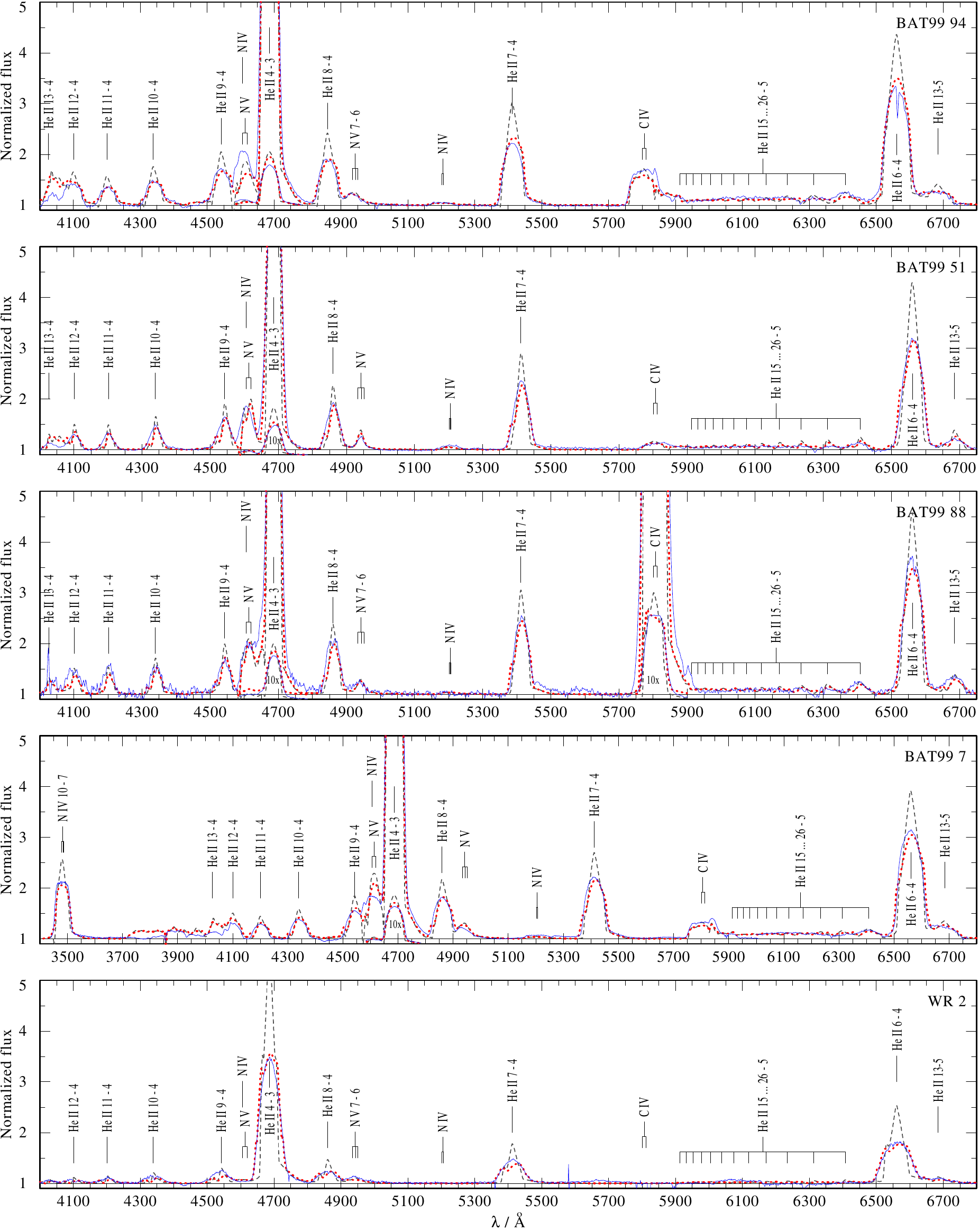}
  \caption{A comparison between synthetic spectra calculated with rotation (dotted red line), without rotation (dashed black line), and observations (solid blue line) 
           for the five stars analyzed.
           The synthetic spectra are convolved with a Gaussian of $\text{FWHM} = 3\AA{}$ to account for instrumental broadening.}
\label{fig:combinedfits}
\end{figure*}

\begin{figure*}[htb]
 \centering
  \includegraphics[width = \hsize]{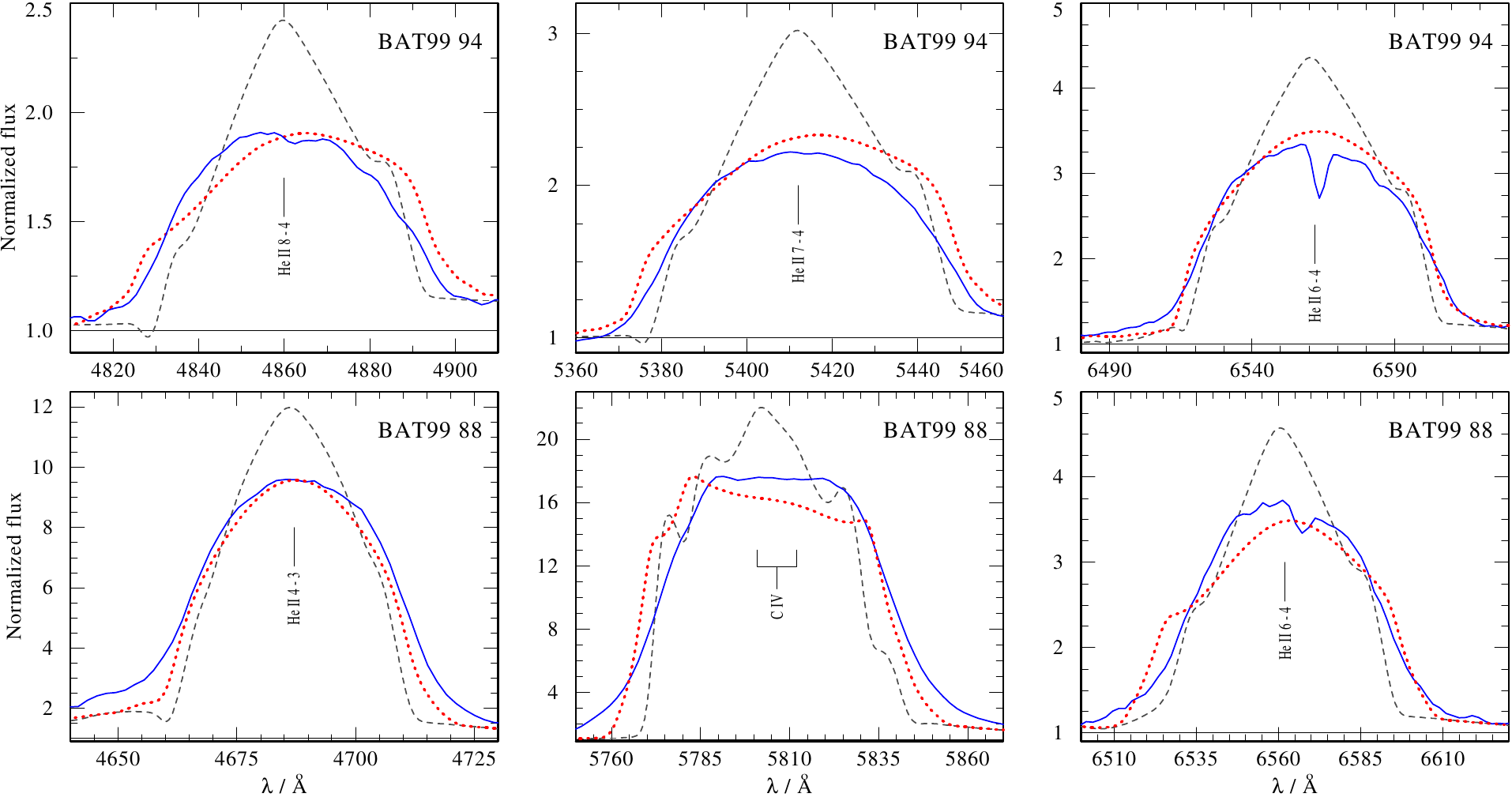}
  \caption{Zoomed-in sections of Fig.\,\ref{fig:combinedfits} with prominent emission lines of the stars BAT99 94 (upper panels) and BAT99 88 (bottom panels). The 
          observed spectrum (solid blue line) is compared to the synthetic spectrum calculated with rotation (dotted red line), and without rotation (dashed black line).}
\label{fig:combinedlines}
\end{figure*}

Table\,\ref{tab:rotstars} compiles the stellar parameters used for the models of the five
stars analyzed here. 
Fig.\,\ref{fig:combinedfits} compares the synthetic spectra  
calculated with rotation (dotted red lines) and without rotation (dashed black lines) 
to the
observations (solid blue lines) 
for the five stars analyzed. For the synthetic spectrum without rotation, we use the same 
parameters from Table\,{\ref{tab:rotstars}.
Fig.\,\ref{fig:combinedlines} zooms on several prominent emission lines of BAT99 94 (upper panels) 
and BAT99 88 (bottom panels) as two illustrative examples which emphasize how rotation helps to reproduce the observations.
As can be seen from Figs.\,\ref{fig:combinedfits} and \ref{fig:combinedlines}, by taking rotation into account,
we manage to obtain a good overall agreement along the spectrum for all five stars analyzed. 
Rotation manages to account for both the broad and round emission profiles as well as the flat-topped ones. 
Fig.\,\ref{fig:combinedlines} illustrates 
the qualitative difference between the line profiles calculated with and without
rotation.  

The observed line widths are not always reached for the individual lines. Furthermore, some synthetic lines exhibit asymmetrical features
of self-absorption which 
do not appear in the observation. This is especially notable in the case of WR 2, though almost all stars exhibit them at certain lines
(Note, for example, the He\,{\sc ii}\,$\lambda6568\,\AA{}$ line of BAT99 88 in Fig.\ \ref{fig:combinedlines}). 
While the models calculated without rotation exhibit such asymmetric features
as well, rotation can either dampen these features
or enhance them. This could partially be due to the discontinuity of the rotational velocity field, to which some 
lines react strongly because their corresponding LFRs intersect the discontinuous domain.
These discrepancies could be avoided if the lines were to be treated individually, because distinct sets of  
parameters can reproduce both their shapes as well as their widths. However, since we aim to treat many lines simultaneously,
a compromise has to be found. 

We mentioned in Sect.\,\ref{sec:introduction} that the round line profiles cannot be reproduced solely by increasing the terminal
velocity $v_\infty$.
As an example, we focus again on the star BAT99 94.   
Fig.\,\ref{fig:vinf_fail} compares the observation (solid blue line) and the calculated synthetic spectrum without rotation (dashed black line) 
for two prominent emission lines of BAT99 94.

For the synthetic spectrum, we use the same stellar parameters as listed in Table\,\ref{tab:rotstars}, 
but choose a larger terminal velocity, so that the line widths
of BAT99 94 are approximately reproduced by the model, and calculate it \emph{without} rotation. 
The suitable value is found to be $v_\infty \approx 2500$\,km/s.
Note that while the line widths are approximately matched, the shape of the observed spectral lines 
appears strikingly different from the synthetic one. We therefore see that larger terminal velocities alone do not help to reproduce
the unique spectra analyzed here.

As mentioned in Sect.\,\ref{subsec:PoWR}, we adopted the value $\beta = 1$ 
for all models in this work, despite the fact that different values of $\beta$ of the 
order of 10
were inferred in the case of a few WR stars \cite[cf.][]{Lepine1999}.
The value $\beta = 1$ was shown by e.g.\ hydrodynamically consistent 
models performed by \cite{Graefener2005} to yield an appropriate approximation of 
the wind velocity field for WR stars. Moreover, $\beta = 1$ enables us to compare our current 
results with previous ones. Still, one could wonder whether a special form 
of the velocity law could explain the round shape of the line profiles.
We have tested 
the effect of varying $\beta$, as well as two-$\beta$ velocity laws, on the synthetic spectra. These tests 
showed that the effect of $\beta$ on most emission line profiles is marginal. 
The obvious reason is that almost all emission lines seen in the optical spectrum 
are formed in regions where $v_r \approx v_\infty$, regardless of the detailed form
of the velocity law (see Fig.\,\ref{fig:tradwl}). 

\begin{figure}[htb]
 \centering
  \includegraphics[width = \hsize]{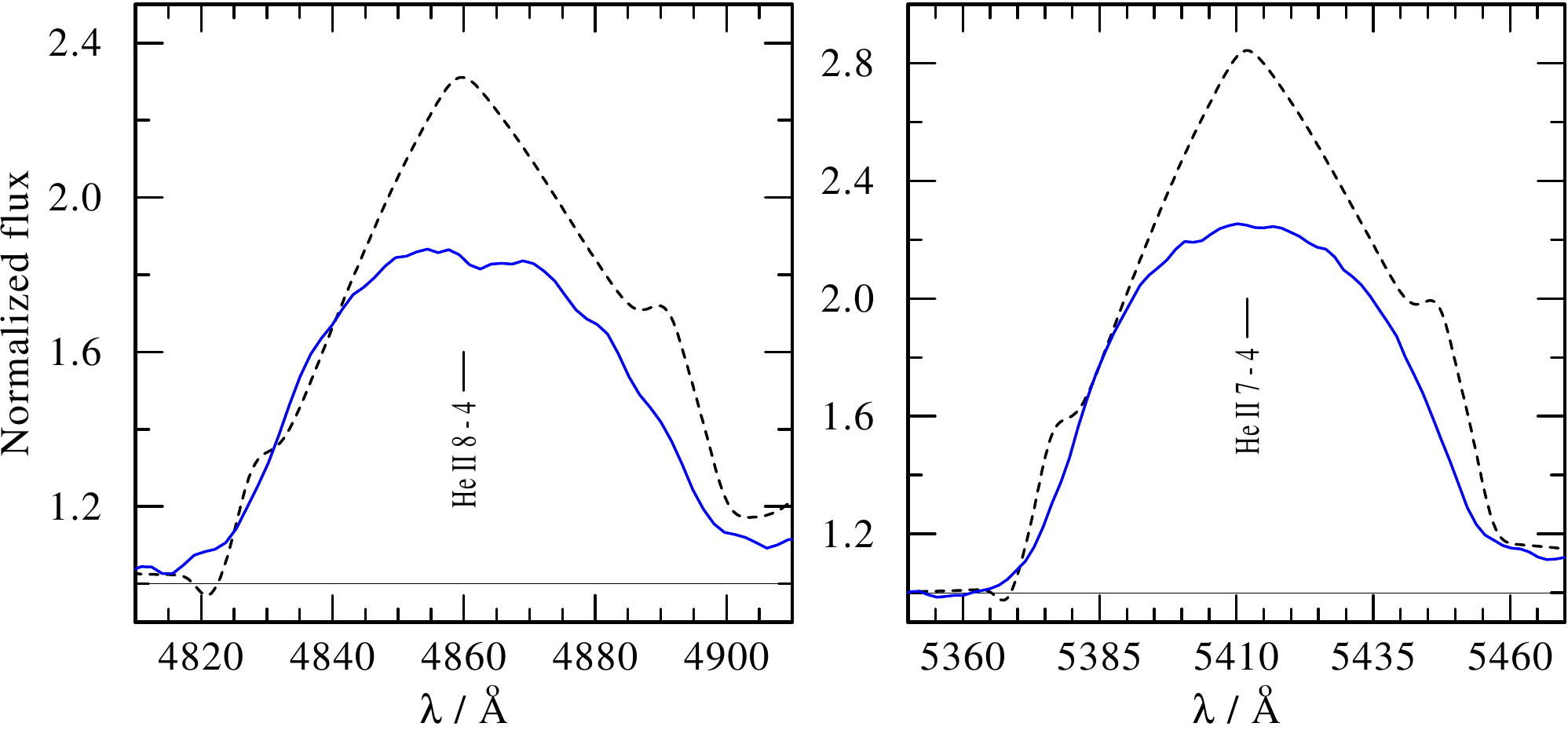}
  \caption{Two He\,{\sc ii} lines belonging to BAT99 94. We compare the observed spectrum (solid blue line) with a synthetic 
          spectrum calculated with the same stellar parameters as in Table\,\ref{tab:rotstars}, but with an increased terminal 
          velocity of $v_\infty = 2500$\,km/s
          and \emph{without} rotation (dashed black line).
Note that the line widths match, but the shape of the synthetic lines 
          is qualitatively very different from the observed ones.}
\label{fig:vinf_fail}
\end{figure}

\subsection{Error estimation}
\label{subsec:Errors}

The uncertainties of the stellar parameters are hard to quantify, since 
they originate in a multitude of sources, e.g.\ deficient atomic data. 
On the basis of past studies, we estimate 
the error in $\log T_*$ to be $0.05$\,dex, and the error in $\log R_\text{t}$ to be $0.1$\,dex. 
The corresponding uncertainty in $\dot{M}$ is $0.15$\,dex. 
For WR 2, which resides in the Galaxy, we estimate an uncertainty in $\log L$ of $0.3$\,dex. 
For the LMC stars, whose distances are known with much better certainty, we estimate an uncertainty in $\log L$ of $0.1$\,dex.
The uncertainty in $v_\infty$ is $20\%$.
As a consequence, $\eta_\text{mom}$ values are given with a factor 2 uncertainty.

The errors of the velocity parameters given in Table\,\ref{tab:rotstars} are estimated on the basis of 
sensitivity tests. 
An impression of the sensitivity of the models to the rotation parameters
may be gained from Fig.\,\ref{fig:partest}. 
We conclude a relative uncertainty in $R_\text{cor}$ of
$30\%$. As for $v_\text{cor} \sin i$, we estimate the 
relative uncertainty to be 
$20\%$. Combining these two estimations, we obtain an uncertainty of $50\%$ in $v_* \sin i$.

\section{Discussion}
\label{sec:discussion}

The results obtained in this work indicate that 
rotation, even with the simplified model now implemented in PoWR, is capable of explaining almost
all peculiarities of the spectra analyzed here: The models reproduce the round emission line profiles and 
flat-topped lines simultaneously.
A good agreement between model and observation is achieved along a large segment
of the optical spectrum.
But are the inferred parameters plausible? We discuss this in the following section.

\subsection{A closer inspection of the stars in the sample}
\label{subsec:inspection}

A glance at the spectral types of the few stars which exhibit round emission lines reveals an interesting fact: All 
round-lined stars are of subtype WNE. Furthermore, all but WR~2 are classified as broad-lined WN stars (suffix b). 
While emission lines of such broad-lined WN stars
indeed tend to be somewhat ``roundish'' compared to the majority of WN stars, the stars in our sample
show a striking difference, as already illustrated in Fig.\ \ref{fig:spitzcomp}. Most importantly, 
the spectra of all other broad-lined WN stars in the Galaxy and the LMC can be reproduced without assuming rotation
\citep[][Hainich et al. in prep.]{Hamann2006}.

The fact that WR~2
is classified as weak-lined (suffix w) could be the result of dilution of its intrinsic spectrum  
due to a luminous OB companion.
It was recently brought to our attention (Moffat, priv. comm.) that 
WR~2 has a faint B-star companion. However, it contributes only $5-10\%$ to the visual continuum, and thus cannot explain
the weak-lined appearance of the spectrum of WR~2.

Since WNE stars are more compact than their WNL counterparts, 
faster rotation in the WNE phase
is plausible. Moreover, the stars in this sample are very compact even compared to typical
WNE subtypes, which implies that these stars approach the end of their life cycle. Indeed, 
a WR star gradually loses its angular momentum due to the stellar wind,
so it is not obvious that an evolved 
WR star should rotate more rapidly than a young one. However, as will 
be discussed below, contraction can win
over the loss of angular momentum, especially at lower metallicities, thus leading 
to an enhancement of the rotational velocity as the WR star evolves.

It is interesting to note that all stars in our sample have a low metallicity. 
WR 2 is located in the metal-poor outskirts of our Galaxy, while the average metallicity in the LMC 
is about 35\% solar \cite[e.g.][]{GrocholskiLMC}. A lower metallicity leads to 
a smaller opacity of the lines which drive the wind, which in turn implies 
a smaller mass-loss rate, a fact which was verified by consistent hydrodynamical models
calculated by \cite{Graefener2008}.
One may therefore conclude
that lower metallicities reduce the loss of angular momentum. Since the outer layers
of the original progenitor star, which rotates approximately as a rigid body, 
possess more specific angular momentum than the inner layers, and since it is these outer layers which are ejected from the star, 
a lower mass-loss rate means faster rotation. A lower metallicity should therefore spur a more rapid rotation
of the continuously contracting star.

One could argue that the WR phase initiates when the
hydrogen has been almost fully removed from the stellar surface, so that, independently of the mass-loss rate,
a certain mass already had to be ejected prior to the WR phase. However,
rotationally induced mixing can also lead to a hydrogen-poor stellar atmosphere, meaning 
that rapidly rotating stars at low metallicities could enter the WR phase with a larger angular momentum \cite[e.g.][]{Yoon2005}. 
More critical, however, is the effect of low metallicities \emph{during} the WR phase. Recall 
that two counteracting mechanisms dictate the rotational velocity of a WR star: the stellar wind, depleting the star
of its angular momentum,
and the contraction, making it rotate faster.
\cite{Meynet2005} calculated 
evolutionary tracks of massive stars with initial masses $M_\text{i}$ larger than $20 M_\odot$ and with initial rotation velocities of $300$\,km/s.
Their models 
clearly predict, at LMC metallicities, and in the mass range $20 M_\odot < M_\text{i} < 60 M_\odot$, 
a significant increase of the rotational velocity as the WR star evolves from a late WN subtype to an early WN subtype.
According to these models,
while the star enters the WR phase with $\sim50$\,km/s as a WNL, its equatorial rotational velocity can increase to a value of up to $\sim200$\,km/s 
in the evolved WNE phase. 
In other words, at LMC metallicities, the contraction of the star seems to dominate over the 
loss of angular momentum due to the stellar wind, causing
it to rotate more rapidly.  
This is in agreement with our finding that only WNE subtypes exhibit round emission line profiles, and that 
most stars in our sample are observed at lower metallicity. 

Note that all stars in our sample have wind efficiencies $\eta_\text{mom}$ larger than unity. For winds which are driven solely by radiation, this 
implies that ``multiple scattering`` occurs. While it is not unusual for $\eta_\text{mom}$ to exceed unity in the case of WNE subtypes,  
a comparison with the wind efficiencies of $\sim 70$ WNE subtypes in the LMC ($\bar{\eta}_\text{mom} \sim 2.0$), 
analyzed by Hainich et al.\,(submitted), reveals that the LMC stars in our sample have 
slightly higher than average wind 
efficiencies. BAT99 7 and BAT99 94, with $\eta_\text{mom} > 5$, are especially striking in this sense, while WR 2 
has a more or less average wind efficiency. 
\cite{Graefener2008} calculated hydrodynamically consistent models for WN stars and obtained, at LMC metallicities,  
a maximum wind efficiency of $\eta_\text{mom} \sim 1.2$. For a WC model with solar metallicity, wind 
efficiencies of up to   
$\sim 2.5$ were reproduced \cite[cf.][]{Graefener2005}. Monte Carlo simulations performed by \cite{Vink2011eta} 
do not yield higher values. 
While such studies verify that multiple scattering occurs in stars approaching the Eddington limit, 
it is still not clear whether multiple scattering alone can explain the very large values of $\eta_\text{mom}$ 
which are inferred from spectral analyses, 
a discrepancy known as the ``momentum problem''.
To solve it, additional wind driving mechanisms were suggested
throughout the years
\cite[cf.][]{Maeder1985, Underhill1987, Poe1989}. Specifically, \cite{Poe1989} 
showed that rapid rotation and magnetic fields can contribute to 
the driving of the wind, a fact we shall return to in Sect.\,\ref{subsec:magfield}. 

After first submission of this paper, our attention has been drawn to a recent study of line profile variations (LPV) in WR~2 performed by Chen\'e et al. (in prep.). 
They detect LPVs of the order of 0.5\% rms, which is smaller than found in other WR stars \cite[cf.][]{Lepine1999}, 
and which they attribute to stochastic clump activity.
A periodicity of the LPVs could not be established. Such periodic variations should occur if co-rotating structures exist which dominate 
over the stochastic clump activity.

\subsection{Are the inferred rotational velocities plausible?}
\label{subsec:veloline} 
All stars analyzed in this work require rapid rotational velocities in order to significantly influence their
corresponding synthetic spectra. But are such velocities conceivable?  

The critical rotation velocity at a 
distance $r \ge R_*$ from 
the center of a star with an effective mass $M_\text{eff}$ is given by
\begin{equation}
 v_\text{crit}(r) = \sqrt{\frac{G M_\text{eff}(r)}{r}},
\end{equation}
where $M_\text{eff} = M_* \left(1 - \Gamma_\text{e}\right)$ is the stellar mass corrected due to 
the radiative force on the electrons.

If we take BAT99 51 as an example, using the parameters from Table\,\ref{tab:rotstars} and an
Eddington factor $\Gamma_\text{e}\left(R_*\right) = 0.23$ as obtained from our model, we find the 
critical velocity at the stellar surface to be
$v_\text{crit}\left(R_*\right) \approx 1050$\,km/s.
The rotational velocity 
at the stellar surface (cf.\ Table\,\ref{tab:rotstars}) is therefore far below the critical velocity: \mbox{$v_* \sin i \approx 0.2\,v_\text{crit}(R_*)$}. 
This implies 
that the star approximately maintains spherical symmetry even for the rotation parameters found in this analysis. 
The same holds for all stars in our sample, with WR 2 having the largest value of  
$v_* \sin i = 0.27\,v_\text{crit}$
at the stellar surface.
However,
due to the co-rotation, the rotational velocity of BAT99 51 grows linearly with $r$ until it reaches the maximum value 
of $v_\text{cor} \sin i = 2000\,$km/s 
at $R_\text{cor}$. The critical velocity of BAT99 51 
at $R_\text{cor}$ is $v_\text{crit}\left(R_\text{cor}\right) \approx 350$\,km/s. We see that the rotational velocity at the co-rotation 
radius exceeds the critical velocity by far. 
This means that the wind should be partially 
driven by centrifugal forces.

Centrifugal forces can, in principle, have two effects on the wind. Firstly, they can affect the mass-loss rate, 
causing it to no longer be isotropic, but rather enhanced at the equator and reduced towards the poles. 
Secondly, they can affect the hydrodynamics of the wind, altering the wind velocity field.

The first effect is not expected to occur with the parameters inferred here. 
According to the theory of radiation driven winds, the mass-loss is not 
affected by forces which act beyond the so-called ``critical point'' \cite[][CAK hereafter]{CAK1975}.
Since the centrifugal
forces become significant only much further away than the critical point, they do not affect the mass-loss
rate. 
The mass flux can thus be assumed to be isotropic.

However, the centrifugal forces implied from our analysis should bear an effect on the wind velocity, 
perhaps even to
the extent that the $\beta$-law (Eq.\,\ref{eq:beta}) no longer holds. The centrifugal force vanishes 
towards the poles, and reaches its maximum at the equator. 
The result is a latitude-dependent wind velocity field, which also induces a latitude-dependent density structure.
While this potentially has important consequences, a proper treatment of this problem would require a 
consistent hydrodynamical modeling of the wind, which is beyond the scope of the current work.

However, the departure from spherical symmetry in the wind may be observable with linear polarimetry, as described
in Sect.\,\ref{subsec:lineeffect}.

\subsection{The line effect}
\label{subsec:lineeffect}

Natural light scattered off free electrons (Thomson scattering) 
is partially linearly polarized.
If the scattering layers of a star are spherically symmetric, the net linear polarization of the continuum radiation 
due to scattering vanishes.
However, if the scattering layers of the star depart from spherical symmetry, 
the linear polarization of the continuum no longer cancels out. To cope with 
the contamination of polarimetric data due to interstellar polarization, 
it is essential to compare the observed continuum polarization to that observed in emission lines.
Emission lines formed far from the stellar surface 
are not expected to show significant polarization because such emission
mainly follows recombination processes and is thus unpolarized. 
A subsequent scattering of these line photons is rare compared to the scattering of continuum photons,  
as can be also deduced from the general weakness of the electron scattering wings 
\cite[][]{Hillier1991_wings, Hamann1998}. Thus, line radiation 
is less polarized compared 
to the continuum, an effect termed the ``line effect''. The line
effect was observed in various WR stars \cite[e.g.][]{Schulte-Ladbeck1991, Harries1998}, and 
attempts for a numerical calculation of the continuum polarization due to scattering were performed 
by e.g.\ \cite{Chandrasekhar1960}
\cite{Brown1977}, \cite{Hillier1991}, and \cite{Hillier1994}. 

%

\cite{Vink2007} analyzed 13 WR stars in the LMC for the line effect, 
out of which two were positively detected. 
Alas, none of the LMC stars
analyzed here for rotation are among this sample. Scholz et al.\,(in prep.)
analyzed polarimetric observations of BAT99 7,
but did not detect line depolarizations larger than $\sim 2\%$. The remaining 
LMC stars analyzed here were so far not studied for linear polarization.
\cite{Harries1998} analyzed 16 Galactic WR stars for the line effect, out of which
four were positively detected.  WR~2  was not among the sampled stars. 
However, \cite{Akras2013} very recently reported a
linear polarization of 3\% in optical broadbands 
of WR~2. 
In contrast,
Chen\'e et al. (in prep.) extensively analyzed polarimetric data of WR~2 and do not find
any linear line depolarization down to the 0.05\% noise level, and attribute 
the 3\% linear polarization detected by \cite{Akras2013} to scattering off grains in 
the interstellar matter.

The non-detection of the line effect in WR~2 challenges its rapid rotation 
suggested in this work.
To allow for a quantitative interpretation of the polarimetric data with respect 
to our rotation model, polarized radiative transfer on top of 
hydrodynamic wind models 
would be needed, which is
far beyond the scope of this work. 
In a preliminary fashion, we can qualitatively discuss 
whether one should expect an observable net continuum 
polarization from co-rotating winds, as considered here.

Any net polarization will be determined mainly by the following three factors: the amount of scatterings 
occurring in each layer, the degree of pole-to-equator
density contrast in each layer, and the proximity of the scattering layers to the stellar surface.

The relevant physical quantity for assessing the relative amount of scatterings in each layer 
is the Thomson optical depth, 
$\tau_\text{th}$. In our model of WR~2, $\tau_\text{th}$ is smaller than 
0.1 for radii $r > 4\,R_*$, 0.3 in the vicinity of $r \sim 2\,R_*$, 
2/3 at $r\sim 1.3\,R_*$, and larger than 1 for radii $r < 1.1\,R_*$. At radii $r > 1.1\,R_*$,
the continuum optical depth $\tau_\text{Ross}$ is almost completely dominated
by $\tau_\text{th}$, meaning that the atmosphere becomes optically thin to continuum at $r \sim 1.3\,R_*$. As 
the number of scatterings is roughly proportional to $\tau_\text{th}$, 
we see that most observed photons have been scattered for the last time in the vicinity 
of $r \sim 1.3\,R_*$. Only few photons are scattered at radii $r > 2\,R_*$.

The 
pole-to-equator density contrast $C = \rho(r,\pi/2)/ \rho(r,0)$ can be expressed in terms of
the mass flux per solid angle $\dot{m}(\theta)$ and the radial wind velocity $v_r(r,\theta)$ at pole and equator
via the mass-continuity equation.
For this purpose, we adopt the standard line-driving 
formalism by \cite{CAK1975}, albeit the CAK theory fails to explain the high mass-loss
rates of WR stars and also does not account for rotation. 

We now follow the arguments by 
\cite{Owocki1998}:
$\dot{m}(\theta)$, determined at the critical point $r_\text{c}$ very close to the stellar surface,
can be expressed in terms of the radiative flux $F(\theta)$, the  
effective gravity $g_\text{eff}(\theta)$, and the force-multiplier parameter $\alpha$. The Von Zeipel
effect \cite[][]{VonZeipel1924} states that $F(\theta) \propto g_\text{eff}(\theta)$, so that one 
eventually obtains \cite[cf.][eq.\ 4]{Owocki1998} $\dot{m} \propto g_\text{eff}(\theta)$, 
independently of $\alpha$. Meanwhile, $g_\text{eff}(\theta) \propto 1 - \Omega \sin \theta$, 
where
$\Omega(r) = \left(v_\text{rot} / v_\text{crit}\right)^2$ denotes the squared ratio
between the rotational and critical velocities at the stellar surface.
Approximating $r_\text{c} \cong R_*$, 
\cite{Owocki1994} obtain $\dot{m}(\pi/2)/ \dot{m}(0)=  1-\Omega$.
For WR~2, which has the largest inferred rotation velocity,
$v_* \sin i \sim 500\,$km/s (cf.\ Tab. \ref{tab:rotstars}), 
and so (ignoring projection effects) $\Omega(R_*) = 0.11$, and 
$\dot{m}(\pi/2) / \dot{m}(0) = 0.89$.

The radial component of the wind velocity at the pole, $v_r(r,0)$, is not affected by the rotation 
and therefore follows the $\beta$-law. At the equator, the centrifugal force in the co-rotating regime
$F_\text{cent} \propto r$ enhances the radial component of the equatorial velocity $v_r(r,\pi/2)$. Estimating the expansion velocity ratio between 
pole and equator, we obtain $v_r(r,0) / v_r(r,\pi/2) \sim 0.57, 0.5$, and $0.37$ at $r\sim 1.3\,R_*, 
2\,R_*$, and $4\,R_*$, respectively. 
$C$ then follows from the continuity equation

\begin{equation}
 C = \frac{\rho(r,\pi/2)}{\rho(r,0)} = \frac{\dot{m}(\pi/2)}{\dot{m}(0)} \frac{v_r(r,0)}{v_r(r,\pi/2)},
\label{eq:conteq}
\end{equation}

Using Eq.\ (\ref{eq:conteq}), we find $C\sim0.5, 0.44,$ and $0.33$
at ${r\sim 1.3\,R_*}, 
2\,R_*,$ and $4\,R_*$, respectively.
Although relatively little scattering occurs in the outermost wind layers,
their asymmetry is likely to cause a net continuum polarization to a certain extent. However, the 
net polarization is very hard to estimate without consistent modeling.
For example,
\citet[][figure~2]{Hillier1994} presented modeling results
of the net polarization for a constant density pole-to-equator contrast of 
$C = 11$ throughout the scattering atmosphere for various total optical depths and inclination
angles. Even in the relatively simplified atmosphere structure he assumed 
(e.g. constant contrast, point source approximation), the net polarization 
is a complicated function of the total optical depth and inclination, and may 
cancel out even with the extreme density contrast adopted by him, depending
on these parameters.
We therefore stress the need for modeling polarized radiative transfer in co-rotating atmospheres. 
Whether an obvious contradiction between the polarimetric data and the rotation hypothesized 
here arises remains to be seen.

\subsection{Are the inferred co-rotation radii plausible?}
\label{subsec:magfield}
The postulated co-rotation of the wind is motivated by the possible existence of strong magnetic 
fields in WR stars. In this section, we estimate the required magnetic field strengths that would
force the wind plasma to co-rotate with the star up to the inferred co-rotation radii.

Many mechanisms were considered
throughout the years which could induce magnetic fields in massive stars. One possibility is that the magnetic 
fields are ``fossil'', i.e.\ remnants from the time of the star formation \cite[e.g.][]{Donati2009}. Another 
possibility is that some dynamo mechanism induces the magnetic field, generated by e.g.\ Tayler-Spruit 
instabilities \cite[cf.][]{Spruit2002}, or by the convective core of the star \cite[][]{Moss1989, MacGregor2003}. 
A third possibility would be a dynamo generated by sub-surface convection zones that are caused due to iron opacity peaks \cite[][]{Cantiello2009}, leading
to the formation of so-called star spots.
Magnetic spots have also been invoked as the likely cause of CIRs 
in hot-star winds. 
Such spiral
patterns that function as footings of magnetic structures may give rise to the discrete absorption components (DACs) that are regularly
observed in O-stars, and also to X-ray emitting shocks in WR winds \cite[cf.][]{Ignace2013}.
The latter mechanism probably does not generate a magnetic field which has a global structure.
The other mechanisms mentioned should generate 
a global magnetic field. 

To obtain an estimation for the magnetic field strengths, we 
follow the work of \cite{Weber1967} and \cite{Blecher1976}. Constraining our discussion to the equator ($\theta = \pi / 2$), we assume 
a global magnetic field of the form $\vec{B}(r) = B_r(r)\,\hat{r} + B_\phi(r)\,\hat{\bphi}$. 
Assuming that the magnetic field lines are frozen in the plasma at the stellar surface, the magnetic field
leaves the stellar surface radially. The $\phi$-component of the magnetic field 
is negligible up to the co-rotation radius, i.e.\  $B_\phi \approx 0$ for $r \lesssim R_\text{cor}$ \cite[cf.][]{Blecher1976}. 
The \emph{Alfv$\acute{\text{e}}$nic Mach number} $M_\text{A}(r)$ 
in Gaussian units is defined by
\begin{equation}
 M_\text{A}^2 = \frac{4 \pi \rho v_r^2}{B_r^2} \approx \frac{4 \pi \rho v_r^2}{B^2}
\label{eq:mach}
\end{equation}
where $\rho(r)$ is the mass density, and $v_r(r)$ is the radial expansion velocity of the wind.
The last equality holds for $r \lesssim R_\text{cor}$.

$M_\text{A}^2$ gives the ratio of the kinetic expansion energy of the 
plasma $\rho v^2 /2$ to the magnetic energy density $B^2 / 8 \pi$.
The \emph{Alfv$\acute{\text{e}}$n radius} $r_\text{A}$ is set by the condition $M_\text{A}\left(r_\text{A}\right) = 1$.
The wind co-rotates approximately up to the 
\emph{Alfv$\acute{\text{e}}$n radius}, i.e.\ $R_\text{cor} \approx r_\text{A}$. 

As a consequence of Maxwell's equation $\nabla \cdot \vec{B} = 0$, the total magnetic flux $B_r r^2$ is
a conserved quantity. Since $B_\phi \approx 0$ up to the Alfv$\acute{\text{e}}$n radius, this means that 
$B \propto \left(r/R_*\right)^{-2}$ for $r \le r_\text{A}$. 
Eliminating $\rho(r)$ from Eq.\,(\ref{eq:mach}) with the mass continuity equation $\dot{M} = 4 \pi \rho v r^2$,
we can express
the magnetic field in terms of the co-rotation radius:

\begin{equation}
 B_*^2 \approx M_\text{A}^2(R_\text{cor}) B_*^2 = 
  \frac{\dot{M} v_\infty}{R_*^2} \left(1- \frac{R_*}{R_\text{cor}}\right) \left(\frac{R_\text{cor}}{R_*}\right)^2
\label{eq:magfield}
\end{equation}
where we used the beta-law (\ref{eq:beta}) with $\beta = 1$ for $v_r(r)$ , and the fact that  
$M_\text{A}(R_\text{cor}) \approx M_\text{A}(r_\text{A}) = 1$

Using Eq.\,(\ref{eq:magfield}), it is possible to estimate the stellar magnetic 
field $B_*$ necessary to confine the matter up to a given
co-rotation radius. A more exact estimate should, however, include the effect of the centrifugal forces
due to rotation.
Table~\ref{tab:mag} compiles the inferred stellar magnetic fields $B_*$, 
derived with Eq.\,(\ref{eq:magfield}).
We also give the magnetic fields $B_{\tau = 2/3}$ at the radius where the Rosseland mean
optical depth $\tau_\text{Ross}$ reaches 2/3, i.e.\ at the photosphere.
The corresponding radii, denoted with $r_{\tau = 2/3}$, are also indicated in Table\,\ref{tab:mag}.
The uncertainty in $\log B_*$ is determined by the uncertainties of the parameters in Eq.\ (\ref{eq:magfield})
and is estimated to be $0.45$\,dex.

\begin{table}[htb]
\caption{Magnetic fields needed to enforce the co-rotation} 
\begin{center}
\small
\begin{tabular}{lrrr}
 \toprule
     Object  & $B_*$      & $r_{\tau = 2/3}$& $B_{\tau = 2/3}$      \\[0.05mm]
             & [kG]         & [$R_*$] &  [kG]          \\ 
 \midrule
  BAT99 7    & $235$& 2.7 & $32$      \\[0.05mm]
  BAT99 51  &  $35$& 1.3  & $21$     \\[0.05mm]
  BAT99 88  & $60$& 1.5   & $26$        \\[0.05mm]
  BAT99 94 & $180$& 2.2  &$38$     \\[0.05mm]
  WR 2     & $40$& 1.7  & $14$    \\
\bottomrule
 \end{tabular}
\label{tab:mag}
\end{center} {\footnotesize
The values of $r_{2/3}$ are extracted from the 
calculated model atmospheres.
$B_{2/3}$ values are calculated using $B \propto r^{-2}$.}
\end{table}

Admittedly, the inferred magnetic fields are very strong. To date, attempts to detect magnetic fields 
in WR stars failed to detect any such fields down to 
several $100$\,G \cite[e.g.][]{Kholtygin2011, Chevrotiere2013}. These studies, however, do not include any stars 
that show round profiles like the stars in our sample.
If these magnetic fields are globally structured, they should be
detectable with circular polarimetry.

Nevertheless, a couple of OB stars were observed to exhibit strong magnetic fields of several kG \cite[see][for a review on 
the topic]{Naze2013}, with
one star even reaching a value of $\sim 20\,$kG \cite[cf.][]{Wade2012}. Since these stars are recognized
as progenitors of WR stars, it is plausible that some WR stars can have photospheric magnetic fields of at least
this order of magnitude. Due to the contraction of the WR 
star relative to its progenitor O~star, the stellar magnetic field $B_*$ becomes
accordingly stronger at the hydrostatic core if the magnetic flux was conserved during the contraction. 
 
Taking this reasoning one step further, magnetic fields of neutron stars, and specifically 
those of magnetars,  often reach values of $10^{14}-10^{15}\,$G \cite[e.g.][]{Esposito_magnetar}. WR stars were 
suggested as direct progenitors of magnetars \cite[cf.][]{Gaensler2005}.
A popular explanation for the remarkably strong magnetic fields of magnetars is   
the contraction of their progenitor stars along with the conservation of the magnetic flux, though this explanation alone might 
be overly simplistic \cite[cf.][]{Spruit2008}. Indeed, comparing a typical magnetar radius of $\sim 10\,$km
\cite[e.g.][]{Deibel2013} with a typical radius $\sim\,R_\odot$ of an early WN star (cf.\ Table \ref{tab:rotstars}), flux 
conservation would imply a 10 orders of magnitude difference between the magnetic field strengths
at the surface of a magnetar and an early WR star, which roughly agrees with the values given in Table~\ref{tab:mag}.
All in all, the magnetic field strengths inferred here
are conveniently situated in a plausible intermediate domain between the progenitor and descendent stars.

This is not the first time such strong magnetic fields are invoked for WR stars.
\cite{Poe1989} 
argued that the winds of some WR stars are 
partially driven by magnetic and centrifugal forces. Recall that all stars analyzed here have a wind efficiency 
$\eta_\text{mom}$ larger than average, which potentially indicates that additional mechanisms are driving the wind 
(cf.\ Section \ref{subsec:stelpam}).
\cite{Poe1989}
take into account the effects of radiation, rotation, and magnetic fields, to conclude that 
global photospheric fields 
of several kG  are 
required to obtain mass-loss rates and terminal velocities observed in some WR stars. 
On the other hand, \cite{Mullan2005} show that magnetic 
fields arising from Tayler-Spruit instabilities \cite[cf.][]{Spruit2002} 
could also reach several kG. 
 
There is, however, another fact we ought to consider. A star with a large mass-loss rate 
and a strong magnetic field 
cannot maintain rapid rotation over a long time. 
This is because the existence of 
magnetic torques efficiently transfer angular momentum from the star to its wind 
\cite[e.g.][]{Hartmann1982}.
\cite{Friend1984} estimated
the ``spin down time'', i.e.\ the time scale over which the angular momentum $J$ is lost, as
\begin{equation}
 \tau_\text{J} = \frac{J}{\mathrm{d}J / \mathrm{d}t} \approx \frac{3}{5}\frac{M_* R_*^2}{\dot{M} r_\text{A}^2}  \approx
  \frac{3}{5}\frac{M_* R_*^2}{\dot{M} R_\text{cor}^2}
\label{eq:spindown}
\end{equation}

We now ask ourselves whether we can expect to actually observe rapidly rotating WR stars. To answer this 
question, we need to compare the spin-down times with a typical lifetime of a WR star.
In order to discuss this important question properly, we restrict the following discussion 
to the LMC stars, since they form the majority in our sample and since they share 
a common environment.

As is readily seen from Eq.\,(\ref{eq:spindown}), the spin-down times strongly depend on the co-rotation radius ($\tau_\text{J} \propto R_\text{cor}^{-2}$). 
At the beginning of its evolution, 
a WR star has hardly contracted relative to its O star progenitor, and therefore has a co-rotation radius of the order of the stellar radius, 
i.e.\ $R_\text{cor} \approx R_*$. The mass-loss rate of a young WR star is smaller than that of an evolved one \cite[][]{Nugis2000, Hamann2006}, and 
its mass is larger. 
Using the parameters from Table\,\ref{tab:rotstars}, we can find a lower limit for the ``initial'' spin-down 
times of the stars analyzed here. These lower limits are found to be of the  
order of $10^6$ years, somewhat longer than the lifetime of a rotating WR star at LMC metallicities, i.e.\ $\sim400\,000$ years \cite[cf.][]{Meynet2005}.
However, as the WR star contracts, the quantity $R_\text{cor} / R_*$ becomes larger. The ``current'' spin-down times
of the evolved WNE subtypes in our sample are of the order of $\sim 5\,000 - 10\,000$ years. As soon as the magnetic 
field enforces a large co-rotation radius, the spin-down times become much smaller. 

It is in the WNE phase that WR stars become relatively compact. According to observations, there are approximately 
as many WNE stars as there are WNL stars, in the Galaxy \cite[see e.g.][]{Hamann2006} as well as in the LMC (e.g.\ Hainich et al.\ submitted). 
This means that the lifetime of 
WNE subtypes is about half of a full WR lifetime, i.e.\ $\sim200\,000$ years. Assuming 
that the spin-down times at the WNE phase is $\sim10\,000$ years, and assuming that most
WR stars at LMC metallicities reach this rapid rotation phase at a certain point, we have, statistically, 
a $10\,000 / 200\,000  = 0.05$ chance of observing them. That is, $\sim5\%$ of the WNE subtypes in the LMC sample should 
exhibit round emission lines,   if rapid rotation and magnetically enforced co-rotation are the causes
for it, and as long as the rotation is not concealed due to an unlucky inclination of the star. 
And indeed, the four stars analyzed here, out of the $\sim 70$ WNE subtypes analyzed by Hainich et al.\ (submitted), make
about $5\%$ of the sample.

If a massive WR star is to form a GRB according to the collapsar model of \cite{Woosley1993_2}, 
the inner core of the star (up to $\sim 3 M_\odot$) should possess a sufficiently large angular momentum 
when collapsing. Are the sampled stars in this analysis potential progenitors of GRBs?
All stars analyzed here are compact WNE subtypes which exhibit broad and round emission lines, i.e.\  
potentially rapidly rotating stars approaching the end of their life cycle, 
thus fulfilling the main requirement of the collapsar model. 
On the other hand, 
it is also clear that not all low-metallicity WR stars end their lives as long-duration GRBs \cite[cf.][]{Woosley2006}. 
GRB progenitors need to be rarer than rapidly rotating WR stars are. 
To become a GRB, the spin-down time should be smaller than the remaining lifetime of the stars.
If the stars in our sample were to explode in the next $5\,000 - 10\,000$ years, they might indeed end their lives as GRBs.
We currently see no good way of determining with certainty whether this is the case, however.

\section{Summary}
\label{sec:summary}

Among the $\sim 180$ WN stars analyzed in the Galaxy and in the LMC, we identified five stars (one galactic, four LMC) whose spectra
exhibit exceptionally broad and round emission line profiles. 
It has been suggested that these profiles might 
indicate rapid rotation \cite[][]{Hamann2006}. 
Motivated by this and by the lack of alternative explanations,
we extended our code to properly handle rotation in expanding atmospheres of hot stars.
For this purpose, we assumed a rotational velocity field which is specified by two parameters:
the co-rotation radius $R_\text{cor}$ and the projected equatorial velocity at the co-rotation radius $v_\text{cor} \sin i$. These parameters, 
together with $v_\infty$, were regarded as free parameters throughout the analysis. 
We conclude:
\begin{itemize}
 \item Rotation helps to reproduce the unique features observed.
 \item If rotation is to explain the peculiar features of these spectra, large co-rotation radii have to be assumed ($R_\text{cor} \gtrsim 10\,R_*$)
 \item The rotational velocities are large ($v_* \sin i \sim 200\,$km/s) but remain significantly smaller than the critical velocity at the stellar surface. In the 
       stellar wind, the rotational velocities exceed the critical velocities
       by far. 
 \item The asymmetry arising in the outer wind layers is likely to induce a net polarization of the continuum, 
       which is not observed in the case of WR~2. However, a conclusive interpretation of the polarimetric data 
       requires a full modeling of polarized radiative transfer in co-rotating atmospheres.
 \item Very large magnetic fields ($B_{\tau = 2/3} \sim 20\,$kG) are necessary to enforce co-rotation up to the inferred radii. If the fields are global, 
       they should be detectable with circular polarimetry.
\end{itemize}

It might well be that the rotation model we applied here is too simplified. Even if the 
assumptions adopted in this work hold fully, the actual 
rotational velocity field, of course, would not consist of two artificially distinct domains, but would rather
continuously switch from one domain to the other, as the magnetic field gradually loses control over the wind dynamics.
Regardless, stellar rotation, as implemented here, was shown to help produce spectra which show a good agreement with observations 
along a large segment
of the optical spectrum for all stars we analyzed.

And yet rotation is not only conjured up to help reproduce the unique spectra analyzed here. There are various arguments 
supporting the existence of rapidly rotating WR stars. 
The progenitors of WR stars are known to exhibit very rapid rotational velocities on average.  
Evolutionary models predict the existence of WR stars with significant rotational velocities. Moreover, rapid internal rotation is a prerequisite
for single-star GRB models.
Effects such as induced mixing and gravitational darkening, which were shown to play a role in the evolution of massive stars, originate in
stellar rotation. Knowing whether WR stars exhibit rapid rotation is thus, in many aspects, important.

So far, no other consistent spectral models were able to reproduce the unique spectra discussed in this work. If not for rotation, 
the origin of these spectra remains a mystery. 

\begin{acknowledgements}
We thank C.\,Foellmi for providing spectroscopic data. We further thank U. R\"uhling
for her inspiring work, and R. Hainich for his extensive spectral analyses
of the LMC WN stars. We sincerely thank A.\,Moffat, whose insights significantly 
deepened the scope of this work.
T. Shenar would like to thank the 
Richard-Winter-Stiftung for their support during 
the writing of this paper.
\end{acknowledgements}

\bibliography{literature}

\end{document}